\documentclass{iau} 
\usepackage{graphicx}

\newcommand{\Msun}{{\rm  M_{\odot}}}
\newcommand{\un}[2]{#1_{\rm #2}}

\title[Multi-wavelength radiative properties of the first galaxies] 
{Galaxy evolution and radiative properties in the early Universe:
multi-wavelength analysis in cosmological simulations}

\author[Arata et al.]   
{Shohei Arata$^1$,
 Hidenobu Yajima$^2$,
 Kentaro Nagamine$^{1,3,4}$,
 Yuexing Li$^{5,6}$
 \and Sadegh Khochfar$^7$}

\affiliation{$^1$Theoretical Astrophysics, Department of Earth and Space Science, Graduate School of Science, Osaka University, Toyonaka, Osaka 560-0043, Japan \\ email: {\tt arata@astro-osaka.jp} \\[\affilskip]
$^2$Center of Computational Sciences University of Tsukuba, Ibaraki 305-8577, Japan \\
$^3$Department of Physics \& Astronomy, University of Nevada, Las Vegas, 4505 S. Maryland Pkwy, Las Vegas, NV 89154-4002, USA\\
$^{4}$Kavli IPMU (WPI), The University of Tokyo, 5-1-5 Kashiwanoha, Kashiwa, Chiba, 277-8583, Japan \\
$^{5}$Department of Astronomy \& Astrophysics, The Pennsylvania State University, 525 Davey Lab, University Park, PA 16802, USA \\
$^{6}$Institute for Gravitation and the Cosmos, The Pennsylvania State University, University Park, PA 16802, USA \\
$^{7}$SUPA, Institute for Astronomy, University of Edinburgh, Royal Observatory, Edinburgh, EH9 3HJ, UK
}

\pubyear{2019}
\volume{352}  
\jname{Uncovering early galaxy evolution in the ALMA and JWST era}
\editors{}

\begin{document}

\maketitle

\begin{abstract}
Recent observations have successfully detected UV or infrared flux from galaxies at the epoch of reionization. However, the origin of their radiative properties has not been fully understood yet. Combining cosmological hydrodynamic simulations and radiative transfer calculations, we present theoretical predictions of multi-wavelength radiative properties of the first galaxies at $z=6-15$. We find that most of the gas and dust are ejected from star-forming regions due to supernova (SN) feedback, which allows UV photons to escape. We show that the peak of SED rapidly shifts between UV and infrared wavelengths on a timescale of 100 Myr due to intermittent star formation and feedback. When dusty gas covers the star-forming regions, the galaxies become bright in the observed-frame sub-millimeter wavelengths. In addition, we find that the escape fraction of ionizing photons also changes between $1-40\,\%$ at $z>10$. The mass fraction of {\sc H\,ii} region changes with the star formation history, resulting in the fluctuations of metal lines and Lyman-alpha line luminosities. In the starbursting phase of galaxies with the halo mass $\sim 10^{11}\,\Msun$ ($10^{12}\,\Msun$), the simulated galaxy has $L_{\rm [O\,III]} \sim 10^{42}\, (10^{43})\, {\rm erg\,s^{-1}}$, which is consistent with the observed star-forming galaxies at $z>7$. Our simulations suggest that deep {\sc [C\,ii]} observation with ALMA can trace the distribution of neutral gas extending over $\sim 20$ physical kpc. We also find that the luminosity ratio $L_{\rm [O\,III]}/L_{\rm [C\,II]}$ decreases with bolometric luminosity due to metal enrichment. 
Our simulations show that the combination of multi-wavelength observations by ALMA and JWST will be able to reveal the multi-phase ISM structure and the transition from starbursting to outflowing phases of high-$z$ galaxies.
\end{abstract}

\firstsection 
\section{Introduction}

Recently, galaxies in the reionization epoch ($z\gtrsim 6$) have been observed in the sub-mm wavelength; for example, dust continuum \cite[(Riechars et al. 2013;  Watson et al. 2015; Laporte et al. 2017; Marron et al. 2018; Hashimoto et al. 2019)]{Riechers13,Watson15,Laporte17}, metal emission lines of {\sc [O\,iii]} 88\,${\rm  \mu m}$ \cite[(Inoue et al. 2016; Carniani et al. 2017; Hashimoto et al. 2018, 2019; Tamura et al. 2019)]{Inoue16,Carniani17,Hashimoto18,Tamura19} and {\sc [C\,ii]} 158\,${\rm  \mu m}$ \cite[(Carniani et al. 2017; Marron et al. 2018, Hashimoto et al. 2019)]{Carniani17,Marron18}.
These galaxies were originally high-$z$ candidates identified by the UV photometric observations \cite[(e.g. Bouwens et al. 2015)]{Bouwens15}.
In particular, recent ALMA observations have allowed us to detect many galaxies at $z>6$ and showed wide variety of radiative properties of the {\it first galaxies}.
However, the origin of the variety have not been understood yet. Therefore, we here investigate the relation between galaxy evolution and radiative properties by combining cosmological simulations and radiative transfer calculations.

Recent state-of-the-art simulations showed that star formation in high-$z$ galaxies occurred intermittently due to SN feedback and gas accretion \cite[(e.g. Kimm \& Cen 2014; Yajima et al. 2017)]{Kimm14,Yajima17}.
In this work, we focus on how the intermittent star formation affects the radiative properties.
Dust is also important for the SED because it absorbs UV photons and re-emits IR radiation.
The dust absorption efficiency changes with the size distribution, composition, and spatial distribution of dusty clouds.
We show that changes of dust distribution driven by SNe induce rapid transitions of radiative properties.

\section{Simulations}

We perform cosmological smoothed particle hydrodynamics (SPH) simulations using the {\sc Gadget-3} code \cite[(Springel 2005)]{Springel05} with the sub-grid models developed in the OWLS project \cite[(Schaye et al. 2010)]{Schaye10} and FiBY project \cite[(Johnson et al. 2013)]{Johnson13}. 
First, we conduct dark-matter only N-body simulations with  $(20\,h^{-1}~{\rm cMpc})^{3}$ and $(100\,h^{-1}~{\rm cMpc})^{3}$ boxes, and identify the most massive haloes at $z=6$ with the friend-of-friend method. Hereafter we call the haloes Halo-11 and Halo-12, respectively. Next we calculate these galaxies with zoom-in initial conditions and hydrodynamics. The details of sub--grid models are described in \cite[Yajima et al. (2017)]{Yajima17}. Table \ref{table:setup} shows the properties of Halo-11 and Halo-12 runs. 

\begin{table}
  \centering
  \begin{tabular}{ccccc}
    \hline
    Halo ID  & $\un{M}{h}~[h^{-1}~\Msun]$  &  $\un{m}{DM}~[h^{-1}~\Msun]$  &  $\un{m}{gas}~[h^{-1}~\Msun]$  &  $\un{\epsilon}{min}~[h^{-1}~{\rm pc}]$\\
    \hline
    Halo-11  & $1.6\times 10^{11}$  & $6.6\times 10^{4}$ & $1.2\times 10^{4}$ & $200$ \\
    Halo-12  & $7.5\times 10^{11}$  & $1.1\times 10^{6}$ & $1.8\times 10^{5}$ & $200$ \\
    \hline
  \end{tabular}
\caption{Parameters of our zoom-in cosmological hydrodynamic simulations: (1) $\un{M}{h}$ is the halo mass at $z = 6$. (2) $\un{m}{DM}$ is the mass of a dark matter particle in the corresponding simulation run. (3) $\un{m}{gas}$ is the initial mass of a gas particle. (4) $\un{\epsilon}{min}$ is the gravitational softening length in comoving units.
 }
\label{table:setup}
\end{table}

For multi-wavelength radiative transfer (RT), we use the {\sc Art$^{2}$} code which calculates the propagation of photon packets considering hydrogen ionization, UV continuum, dust emission and Lyman-$\alpha$ line \cite[(Li et al. 2008; Yajima et al. 2012)]{Li08,Yajima12}. This code utilizes the adaptive refinement grids following the simulated gas distribution, which enables the RT computation with the minimum physical scale of $\sim 2.7\,h^{-1}~{\rm pc}$ for Halo-11 at $z\sim 6$. The dust mass in each cell is proportional to the gas metallicity.  We focus on how the RT results depend on the dust distribution. More details are described in \cite[Arata et al. (2019)]{Arata19}.

We also analyze metal line emissions of {\sc [O\,iii]} $88\,{\rm \mu m}$ and {\sc [C\,ii]} $158\,{\rm \mu m}$ using ionization structure of hydrogen estimated by the RT calculation. We first calculate O$^{2+}$ and C$^{+}$ abundances in each cell assuming ionization equilibrium under the stellar radiation field. Next we calculate the rate equations of energy levels, and obtain the metal line luminosities. More details of the model will be described in Arata et al. (in prep.).

\begin{figure}
\begin{center}
\includegraphics[width=\columnwidth]{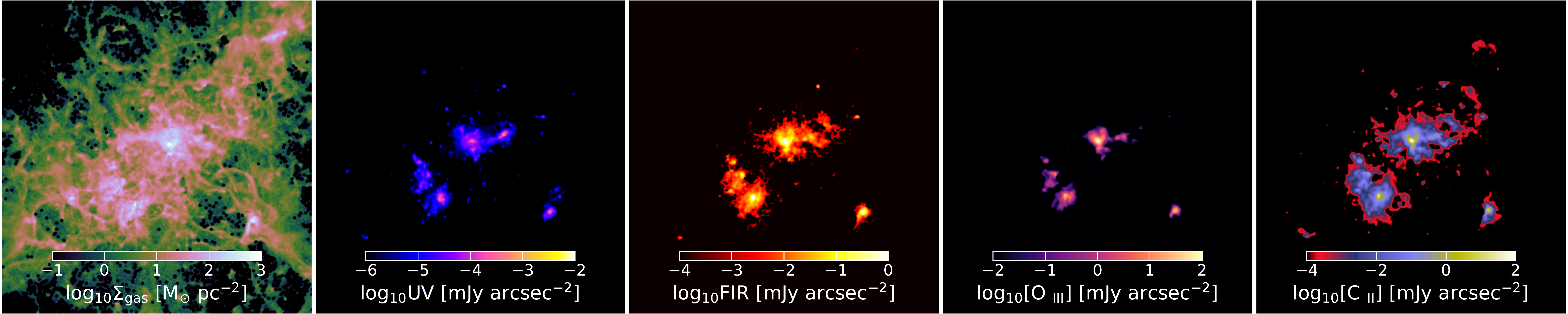}
\caption{Maps of Halo-11 at $z=6.0$. From left to right panels: gas surface density, surface brightness of UV, FIR, {\sc [O\,iii]} 88\,${\rm \mu m}$, and {\sc [C\,ii]} 158\,${\rm \mu m}$ in the rest-frame, respectively. The field of view is $\sim 51\,{\rm physical\ kpc}$. The pixel size is $\sim 0.07\,{\rm arcsec}$.
}
\label{fig:map}
\end{center}
\end{figure}

Figure\,\ref{fig:map} presents the images of Halo-11 at $z=6.0$. The rest-UV surface brightness traces the stellar distribution, and rest-FIR traces dust distribution. The galaxy has a clumpy structure, and each clump changes its color with intermittent star formation (see next section), which results in a large spatial offset of $\sim 1.4\,{\rm arcsec}$ between the brightest pixels of UV and FIR wavelength. 
The {\sc [O\,iii]}  and {\sc [C\,ii]} maps trace the gas distributions of ionized and neutral phases, respectively. Future deep {\sc [C\,ii]} observations ($\gtrsim 10^{-4}\,{\rm mJy~arcsec^{-2}}$) may reveal an extended neutral gas over $\sim 20\,{\rm kpc}$ \cite[(Fujimoto et al. 2019)]{Fujimoto19}.

\section{Rapid transition between UV and IR bright phases}

\begin{figure}
\begin{center}
\includegraphics[width=0.9\columnwidth]{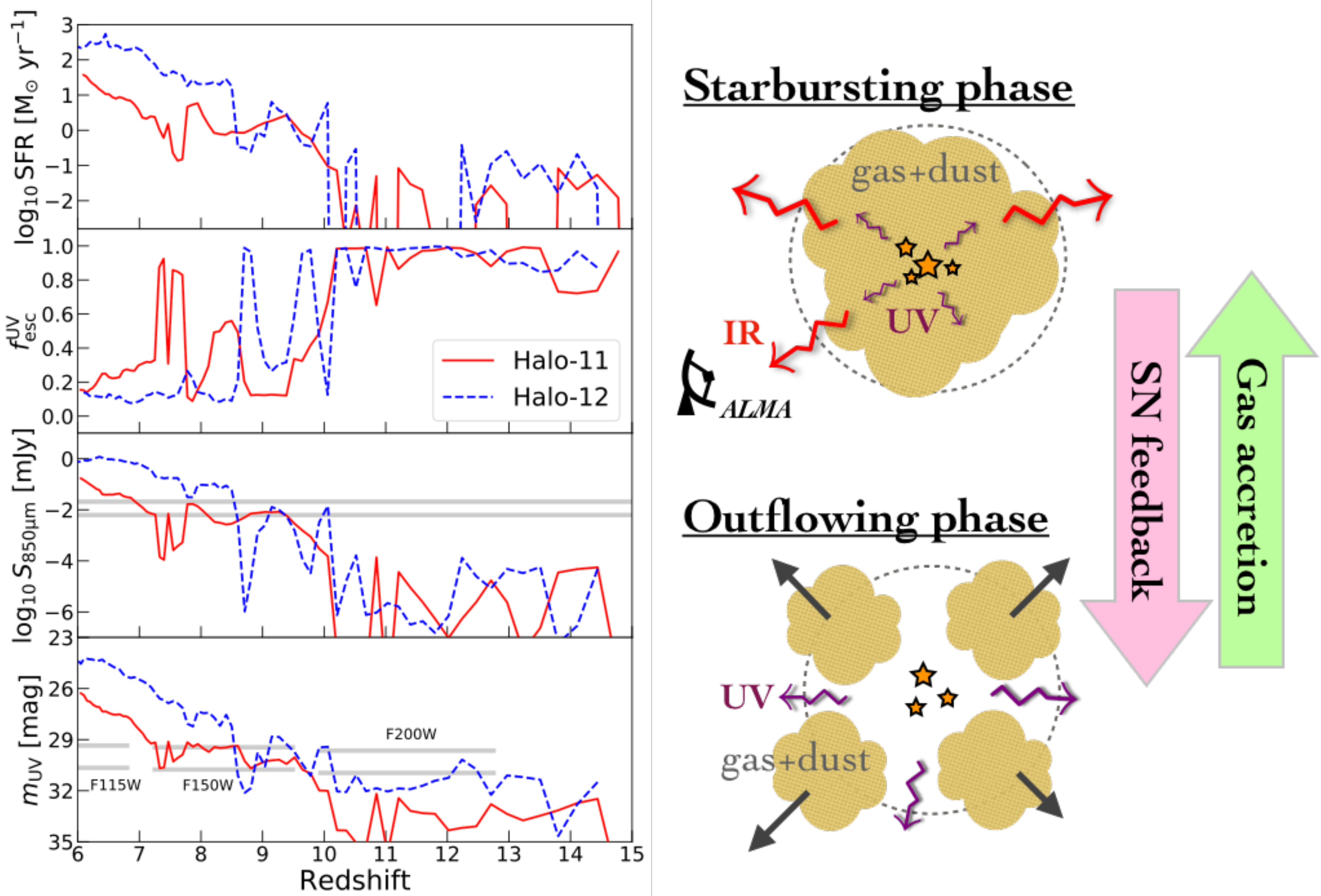}
\caption{{\bf Left}: Redshift evolution of SFR, escape fraction of UV photons, sub-mm flux and apparent UV magnitude in Halo-11 and Halo-12 runs. The gray horizontal lines in the third and forth panels show 3$\sigma$ and 10$\sigma$ detection thresholds for fully operated ALMA and JWST with 10-hour integration. 
{\bf Right}: Schematic picture of relation between galaxy evolution and radiative properties. First galaxies rapidly make transitions between UV and IR bright phases due to intermittent star formation.
}
\label{fig:evolution}
\end{center}
\end{figure}

Figure~\ref{fig:evolution} shows the redshift evolution of SFR, escape fraction of UV photons, sub-mm flux and apparent UV magnitude for Halo-11 and Halo-12 at $z=6-15$.
As described in \cite[Yajima et al. (2017)]{Yajima17}, star formation in the first galaxies occurs intermittently because of gas accretion and SN feedback processes \cite[(see also, Kimm \& Cen 2014)]{Kimm14}. 
In star-bursting phase,  central dusty gas efficiently absorbs UV photons, while most of UV photons can escape from the halo in the outflowing phase, which results in the fluctuation of escape fraction ($f_{\rm esc}=0.2-0.8$ at $z<10$) as presented in the second panel (see also right picture).
The timescale of fluctuation is $\sim 100\,{\rm Myr}$, which corresponds to the free-fall time of the halo.  
The third panel shows that the sub-mm flux of Halo-11 (Halo-12) becomes $\sim 10^{-2}\,{\rm mJy}$ ($\sim 1\,{\rm mJy}$) at $z\sim 7$. Dust mass is $\sim 5\times 10^{6}\,\Msun$ ($\sim 8\times 10^{7}\,\Msun$). Our simulations are consistent with recent ALMA observations: \cite[Watson et al. (2015)]{Watson15} observed dust emission of $0.61\pm 0.12\,{\rm mJy}$ for a galaxy at $z\approx 7.5$, and the estimated dust mass was $4\times 10^{7}\,\Msun$, which are intermediate values of Halo-11 and Halo-12.

To investigate the sub-mm observability of high-$z$ galaxies statistically, we also analyze the radiative properties of all  satellite galaxies in Halo-11 and Halo-12 zoom-in regions. 
We find that the observability of $M_{\rm h} \gtrsim 10^{11}\,\Msun$ galaxies at $z<7$ exceeds $50\,\%$ for a detection limit of $0.1\,{\rm mJy}$.
In addition, we predict that the number density of sub-mm sources observed by future deep survey ($> 10^{-2}\,{\rm mJy}$) will be $10^{-2}\,{\rm cMpc^{-3}}$, which agrees with present UV observations \cite[(see Arata et al. 2019 for details)]{Arata19}.

\section{Star-bursting {\sc [O\,iii]} emitters and quiescent {\sc [C\,ii]} emitters}

The {\sc [O\,iii]} line is emitted only in star-bursting phase because O$^{2+}$ ions exist in {\sc H\,ii} regions formed by massive stars. Thus intermittent star formation results in luminosity fluctuation in the range of $\sim 10^{39}-10^{42}\,{\rm erg~s^{-1}}$ at $z<10$ for Halo-11.
Meanwhile, the {\sc [C\,ii]} line is continuously emitted from neutral regions even in the outflowing phase.

\begin{figure}
\begin{center}
\includegraphics[width=0.9\columnwidth]{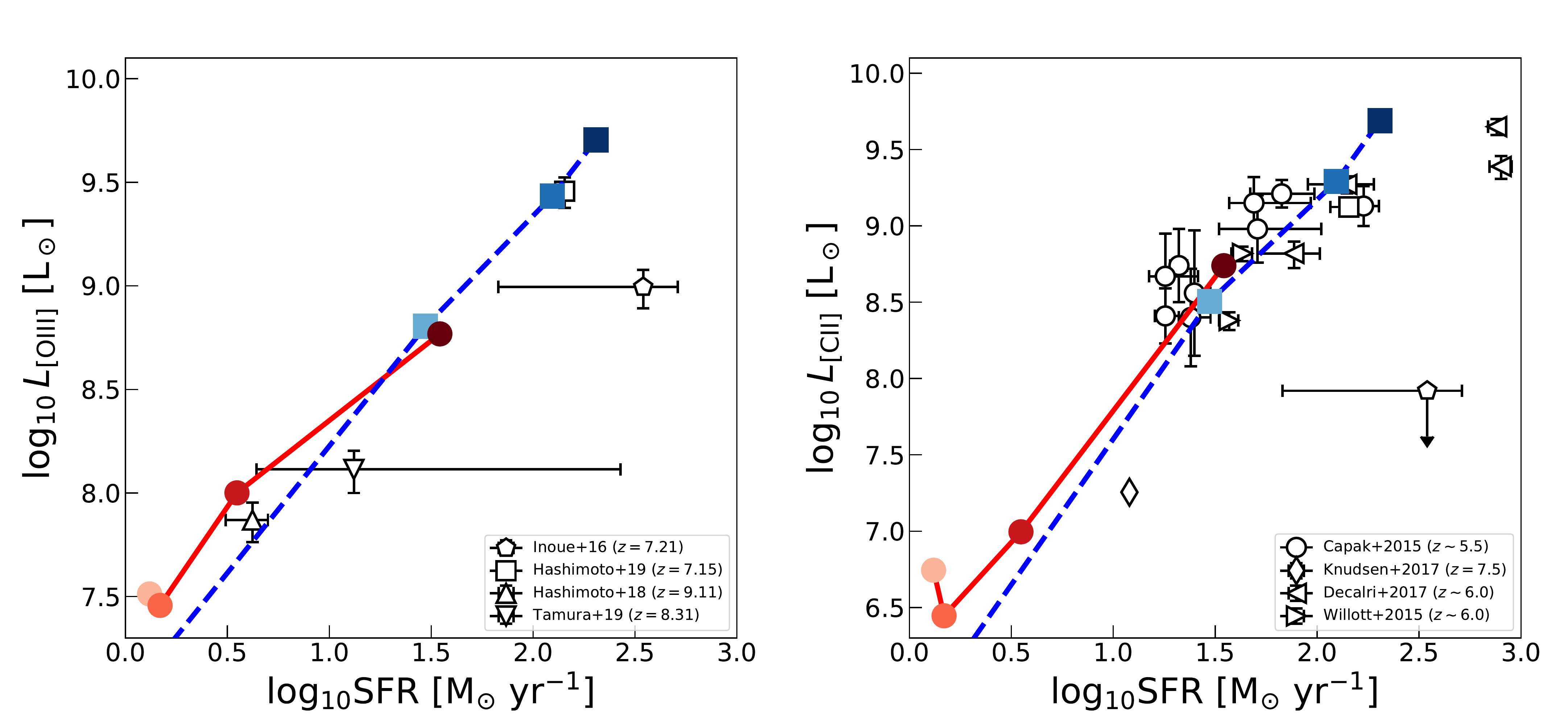}
\caption{Relation between SFR and metal line luminosities of {\sc [O\,iii]} 88\,${\rm \mu m}$ (left) and {\sc [C\,ii]} 158\,${\rm \mu m}$ (right). Red solid line and circles represent evolution of Halo-11 at $z=6,~7,~8,~9$ (darker to lighter). 
Blue dashed line and squares are for Halo-12. Open symbols represent observations.
}
\label{fig:metal}
\end{center}
\end{figure}

Figure~\ref{fig:metal} presents the relation between SFR and metal line luminosities of Halo-11 and Halo-12. 
Our simulations well reproduce the observations of both {\sc [O\,iii]} and {\sc [C\,ii]}  lines.
Note that, however, observational points may shift to higher SFR because actual dust temperature might be higher than the assumed one \cite[(Arata et al. 2019; Ma et al. 2019)]{Arata19,Ma19}.
We will examine the physical properties of very {\sc [C\,ii]}-faint galaxies \cite[(Inoue et al. 2016; Knudsen et al. 2017)]{Inoue16,Knudsen17} in our future paper.

\begin{figure}
\begin{center}
\includegraphics[width=0.8\columnwidth]{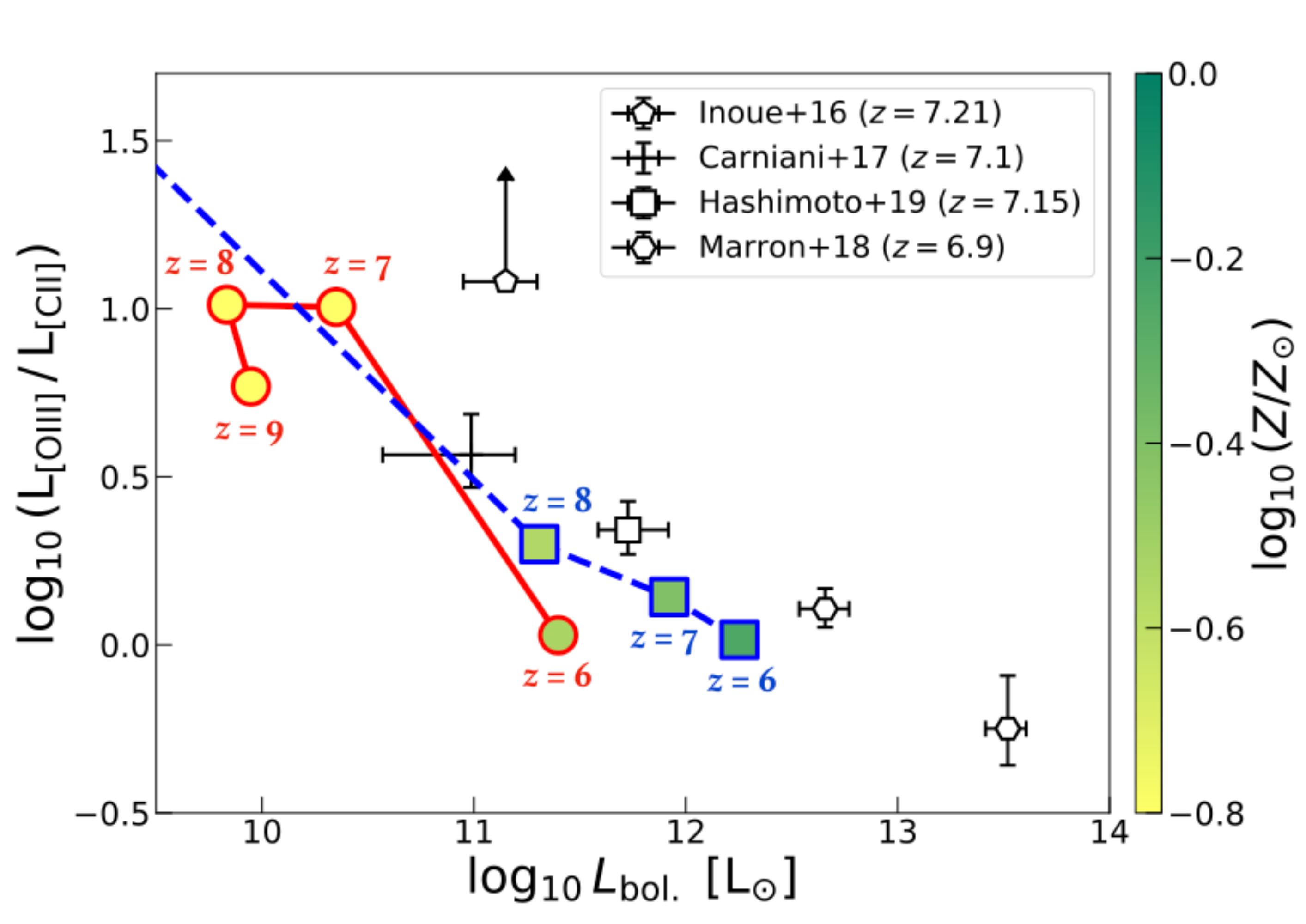}
\caption{Relation between luminosity ratio $L_{\rm [O\,III]}/L_{\rm [C\,II]}$ and the bolometric luminosity ($L_{\rm bol} = L_{\rm UV}+L_{\rm IR}$). Red solid line and circles represent the evolution of Halo-11, and blue dashed line and squares are for Halo-12. The color indicates gas metallicity as shown in the color bar. We find that the origin of negative correlation is metal enrichment: {\sc [O\,iii]} luminosity decreases due to dust absorption of ionizing photons, and $L_{\rm bol}$ increases due to metal cooling. Open symbols represent observed galaxies at $z\sim 7$.
}
\label{fig:ratio}
\end{center}
\end{figure}

Finally, we focus on the luminosity ratio of $L_{\rm [O\,III]}/L_{\rm [C\,II]}$. \cite[Hashimoto et al. (2019)]{Hashimoto19} suggested a negative correlation between  $L_{\rm [O\,III]}/L_{\rm [C\,II]}$ and the bolometric luminosity measured from UV and IR fluxes.
Figure~\ref{fig:ratio} shows the evolution of the ratio for Halo-11 (red circles) and Halo-12 (blue squares) at $z=6-9$.
We find that the ratio decreases about one order of magnitude as gas metallicity increases from sub-solar to solar metallicity. When metal enrichment occurs, dust efficiently absorbs ionizing photons, which results in the shrinkage of {\sc H\,ii} regions. Thus {\sc [O\,iii]} emission is suppressed.
Future deep {\sc [O\,iii]} and {\sc [C\,ii]} observations will reveal the connection between multi-phase ISM structure and the radiative properties for high-$z$ galaxies. 

\end{document}